\documentclass[preprint,onecolumn,nofootinbib]{revtex4}
\pdfoutput=1
\usepackage[colorlinks=true,linkcolor=blue,urlcolor=blue,filecolor=black,citecolor=red,pdfstartview=FitV,pdftitle={},pdfsubject={},pdfkeywords={},pdfpagemode=None,bookmarksopen=true]{hyperref}
\usepackage{graphicx}
\usepackage{amsmath}
\usepackage{amsfonts}
\usepackage{amssymb,ulem}
\usepackage{color}%
\usepackage{dcolumn}
\usepackage{MnSymbol,wasysym}
\usepackage{braket}
\usepackage{eurosym}
\usepackage{calrsfs}
\usepackage[usenames,dvipsnames,svgnames]{xcolor}

\newcommand{\RNum}[1]{\uppercase\expandafter{\romannumeral #1\relax}}

\begin{document}
\baselineskip=0.5 cm
\title{Probes of holographic thermalization in a simple model with momentum relaxation}
\author{Yong-Zhuang Li$^{1,2}$}
\email{liyongzhuang@just.edu.cn}
\author{Xiao-Mei Kuang$^{2,3}$}
\email{xmeikuang@yzu.edu.cn}

\affiliation{\small{$^1$~School of Science, Jiangsu University of Science and Technology, Zhenjiang 212003, China}}
\affiliation{\small{$^2$~Center for Gravitation and Cosmology, College of Physical Science and Technology, Yangzhou University, Yangzhou 225009, China}}
\affiliation{\small{$^3$~School of Aeronautics and Astronautics, Shanghai Jiao Tong University, Shanghai 200240, China}}

\begin{abstract}
\vspace*{0.6cm}
\baselineskip=0.5 cm
From the viewpoint of AdS/CFT correspondence, we investigate the holographic thermalization process in a four dimensional Einstein-Maxwell-axions gravity theory, which is considered as a simple bulk theory dual to a boundary theory with momentum relaxation.
We probe the thermalization process using the equal time two-point functions and the entanglement entropy with the circle profile. We analyze the effects of momentum relaxation on the process in details and results show that the momentum relaxation gives longer thermalization time, which means it suppresses the holographic
thermalization process. This matches the properties of the quasi-normal frequencies for the bulk
fluctuations which the frequency violates from zero mode more profoundly for stronger momentum relaxation. We claim that is reasonable because the decay of the bulk fluctuations holographically describes the approach to thermal equilibrium in the dual theory.
\end{abstract}

\maketitle

\section{Introduction}
The AdS/CFT correspondence \cite{Maldacena:1997re,Gubser:1998bc,Witten:1998qj} provides deep insights into the mapping between the
strongly coupled field theory and the dynamics of a classical gravitational theory in the bulk. Plenty of applications have been made as this correspondence has been treated as a powerful tool to deal with the strongly coupled systems. Among them, the holographic thermalization process, which constructs a proper model in the
bulk gravity to investigate the thermalization process of  quark-gluon plasma (QGP) \cite{Danielsson:1999zt},  has attracted considerable attentions in the theoretical study.
Two main reasons may contribute to the situation. On one hand, the perturbative quantum chromodynamics (QCD) breaks down during the far from equilibrium process of QGP formation. Such that the study of non-equilibrium dynamics in strongly coupled theories are involved in Relativistic Heavy Ion Collider (RHIC) \cite{RHIC1,RHIC2}. On the other hand, the holographic study of hot QCD matter with  equilibrium and near-equilibrium aspects has been well addressed  in \cite{CasalderreySolana:2011us} and further use of the powerful AdS/CFT correspondence to study the thermalization of strongly coupled plasma is a natural clue. Previous attempts were made to describe the thermalization process holographically dual to the formation of  black hole via gravitational collapse in AdS space \cite{Danielsson:1999fa,Janik:2006gp,Janik:2006ft,Chesler:2009cy,Garfinkle:2011hm,Garfinkle:2011tc}.

Later, a simpler model for holographic thermalization was proposed in \cite{Balasubramanian:2010ce,Balasubramanian:2011ur},
which  captures many critical features of the thermalization process. In the proposal, the thermalization process is dual to the collapse of a thin shell of matter described by an Vaidya-AdS metric leading to a thermal equilibrium configuration given by a Schwarzschild-AdS black hole. The time evolution of non-local thermalization probes of the boundary field theory are then dual to the related descriptions in terms of geometric quantities. The authors studied the equal time two-point correlation functions of local gauge invariant operators,
expectation values of Wilson loop operators, and entanglement entropy which are dual to
 minimal lengths, areas, and volumes in AdS space, respectively. Especially, they found that the UV modes thermalize first while IR modes thermalize later which are different from that in perturbative approaches \cite{Baier:2000sb}. And
the thermalization time scales typically as $\sim \ell$, where $\ell$ is the characteristic length of the
probe.  This proposal was soon extended to describe  more real RHIC processes by including  the effect of a non-vanishing
chemical potential \cite{Caceres:2012em,Galante:2012pv}. In this case, the charged matters was involved and the thermal equilibrium configuration was given by a Reissner-Nordstr$\ddot{o}$m(RN)
AdS black hole.  They found that the thermalization time for renormalized geodesic lengths and minimal area surfaces is longer for the larger charge.
Considerable efforts has then been made in  further investigations
on this directs, see for example \cite{Chesler:2011ds,Wu:2012rib,vanderSchee:2012qj,Liu:2013iza,Sin:2013yha,Ageev:2014mma,
Giordano:2014kya,Dey:2015poa,Craps:2015upq,Lin:2015acg,Grozdanov:2016vgg,Zhang:2015npb,
Zeng:2013mca,Li:2013cja,Zhang:2014cga,Arefeva:2013kvb,Fonda:2014ula,Roychowdhury:2016wca,
Alishahiha:2014cwa,Camilo:2014npa,Ghaffarnejad:2018aui,Zeng:2014xpa,Ghaffarnejad:2018vry,
Ageev:2016gtl,Bai:2014tla,Mozaffara:2016iwm,Arefeva:2017pho,Ling:2019tbi,Attems:2017jhep,
Attems:2018prl,Atashi:2016epjc} and therein.

In this paper, we shall study the holographic thermalization via Einstein-Maxwell gravity coupled with two linear spacial-dependent  scalar fields in the bulk. It was pointed out in \cite{Andrade:2013gsa} that the scalar fields in the bulk source a spatially dependent field theory with momentum relaxation and the linear coefficient of the scalar fields  describes the strength of the momentum relaxation. This means that our thermalization process will involve in  momentum relaxation, which is more closer to that may happen in heavy ion collisions as it has been holographically described via boost-invariant plasma and hydrodynamic features in fully inhomogeneous case \cite{Chesler:2009cy,Heller:2011ju,Balasubramanian:2013oga}.

Our aim is to study the effect of momentum relaxation during the thermalization process  from a simple bulk gravity without translational invariance.
We will mainly focus on the  equal time two-point correlation functions of local gauge invariant operators and entanglement entropy which are dual to minimal lengths and volumes in AdS space, and study the effect of momentum relaxation on the two observables. The effect of  momentum relaxation on  the time dependent optical conductivity \cite{Bagrov:2017tqn} and the equilibrium chiral magnetic effect \cite{Fernandez-Pendas:2019rkh}  in the thermalization process of this model has been investigated. It is noticed that the  thermalization process in massive gravity has been studied in \cite{Hu:2016mym}. The authors mainly studied
the effect of dissipation of momentum dually introduced by the massive graviton in the bulk gravity on the equal time two-point correlation functions, and it was found
 that the dissipation of momentum deduced in massive gravity  shortens the holographic thermalization process.
 Even so, our study of  momentum relaxation in  the thermalization process is still deserved from two aspects. One is that both the shell collapsing into the Vaidya-AdS black brane but the
 thermal equilibrium configuration in our study are different from that in massive gravity. The other is that here the mechanism of dissipation of momentum is dual to  breaking of
 translational invariance in the present bulk while it is dual to the breaking of diffeomorphism symmetry in massive gravity \cite{Vegh:2013sk}. Moreover, we will include non-vanishing chemical potential in the study.

The paper is organized as follows. In section \ref{sec:background}, we present the generalized Vaidya-AdS  black brane in Einstein-Maxwell-axions theory. We analyze the effect of momentum relaxation on the holographic thermalization process via numerically computing the two observables, i.e., equal time two-point correlation functions  and entanglement entropy in section \ref{sec:ads2point} and section
\ref{sec:adsHEE}, respectively. The last section is our conclusion and discussion.

\section{Vaidya AdS black branes  in  Einstein-Maxwell-axions theory}\label{sec:background}
We consider the AdS black branes in Einstein-Maxwell-axions gravity theory proposed  in \cite{Andrade:2013gsa}. The action of the four dimensional theory  was given by
\begin{equation}
S=\frac{1}{16\pi }\int \! d^4x \sqrt{-g} \left(R+\frac{6}{l^2}-\frac{1}{4}F_{\mu\nu}F^{\mu\nu}-\frac{1}{2}\sum_{I=1}^{2}(\partial\psi_I)^2\right)\ ,
\label{eq:action}
\end{equation}
By setting the scalar fields to linearly depend on the two dimensional spatial coordinates $x^a$, i.e., $\psi_I=\beta\delta_{Ia}x^a$
\footnote{In general, the linear combination form of the scalar fields are $\psi_I=\beta_{Ia}x^a$. Then defining a
constant $\beta^2\equiv \frac{1}{2}(\sum_{a=1}^{2}\sum_{I=1}^{2}\beta_{Ia}{\beta_{Ia}})$ with the coefficients
satisfying the condition $\sum_{I=1}^{2}\beta_{Ia}{\beta_{Ib}}=\beta^2\delta_{ab}$, we will obtain the same black hole solution.
Since there is rotational symmetry on the $x^a$ space, we can choose $\beta_{Ia}=\beta\delta_{Ia}$ without loss of generality.}, the action admits the charged black brane solution
 \begin{eqnarray}\label{eq-metric}
ds^2&=&-r^{2}f(r)dt^2+\frac{1}{r^{2}f(r)}dr^2+r^2(dx_1^2+dx_2^2),\nonumber \\
f(r)&=&\frac{1}{l^2}-\frac{\beta^2}{2\, r^{2}}-\frac{m_0}{r^3}+\frac{q_{0}^{2}}{r^{4}},\nonumber\\
 A&=&A_t(r) dt, ~~~A_t=\left(1-\frac{r_h}{r}\right)\frac{2q_0}{r_h}
 \end{eqnarray}
where the horizon $r_h$ satisfies $f(r_h)=0$; $l$ describes the radius of AdS spacetime, and for  simplicity we will set $l=1$. $m_0$ and $q_0$ are the mass and charge of the black brane, with the relation given by
\begin{equation}
1-\frac{\beta^{2}}{2r_{h}^{2}}-\frac{m_0}{r_{h}^{3}}+\frac{q_{0}^{2}}{r_{h}^{4}}=0.
\label{massandcharge}
\end{equation}
It is worthwhile to mention that the scalar fields in the bulk source a spatially dependent boundary field theory with momentum relaxation, which is dual to a homogeneous and isotropic black brane \eqref{eq-metric}. The linear coefficient $\beta$ of the scalar fields is usually considered to describe the strength of the momentum relaxation in the dual boundary theory \cite{Andrade:2013gsa}. A general action with axions terms and the holography has been studied in \cite{Alberte:2015jhep}. We note  that the extended thermodynamics of the black brane has also been studied in \cite{Fang:2017nse,Cisterna:2018jqg}. The Hawking temperature of the black brane reads
\begin{eqnarray}\label{adstemperature}
T=\frac{1}{4 \pi}\frac{d\left(r^{2}f(r)\right)}{dr}{\mid_{r_h}}.
\end{eqnarray}
Such a temperature is treated as the temperature of the dual boundary field. The condition $T=0$ then restricts the maximal value of charge parameter $q_0$ if we fix $m_0$ and $r_h$. Besides, the chemical potential of the dual field on the boundary can be modeled by the non-zero electric field and is given by \cite{Caceres:2012em}
\begin{eqnarray}\label{adschemical}
\mu=\lim_{r\rightarrow \infty}A_{t}(r)=\frac{2q_0}{r_h}.
\end{eqnarray}

With properly chosen coordinate transformation, the above black hole brane $\eqref{eq-metric}$ can be represented as in the Eddington-Finkelstein coordinates,
\begin{eqnarray}
ds^{2}&=&\frac{1}{z^2}\left[-f(z)dv^2-2dvdz+dx^{a}dx^{a}\right],\label{metriads1}\\
f(z)&=&1-\frac{1}{2}\beta^{2}z^{2}-m_{0}z^{3}+q_{0}^{2}z^{4},\label{metriads2}
\end{eqnarray}
with
\begin{eqnarray}\label{coortrans}
&&dv=dt-\frac{1}{f(z)}dz \qquad \mathrm{and}\qquad z=\frac{1}{r}.
\end{eqnarray}
We note that the coordinates $v$ and $t$ coincide on the boundary.

In the framework of  AdS/CFT duality, the rapid injection of energy followed by the
thermalization process in the boundary theory corresponds to the collapse of a black brane or a falling thin shell of dust in the AdS spacetime\footnote{Note that metric (\ref{metriads1}) is an ingoing metric in Vaidya-AdS spacetime, which means that the shell will ``fall'' from $z=\infty$ to $z=0$. Correspondingly, in AdS spacetime, the shell is outgoing till it reaches the equilibrium at $r=\infty$.}. Thus, in order to  holographically describe the thermalization process, one usually frees the mass and charge  parameter
as smooth  functions of $v$ as \cite{Balasubramanian:2011ur,Galante:2012pv}
\begin{eqnarray}
m(v)&=&\frac{m_0}{2}\left[1+\text{tanh}\left(\frac{v}{v_0}\right)\right],\label{dynamasspara}\\
q(v)&=&\frac{q_0}{2}\left[1+\text{tanh}\left(\frac{v}{v_0}\right)\right],\label{dynachargepara}
\end{eqnarray}
where $v_0$ represents the finite thickness of the falling charged dust shell and  the relation between $m_0$ and $q_0$ is still given by \eqref{massandcharge}. Then the related Vaidya AdS black brane  is\footnote{It is noticed that the quenches we consider here are homogeneous in spatial directions because the metric in the bulk is homogeneous. However, as we mentioned before that the boundary theory has momentum relaxation sourced by the spacial dependent scalar fields in the bulk.}
\begin{eqnarray}
ds^{2}&=&\frac{1}{z^2}\left[-f(v, z)dv^2-2dvdz+dx^{a}dx^{a}\right], \label{Vaidyametriads1}\\
f(v, z)&=&1-\frac{1}{2}\beta^{2}z^{2}-m(v)z^{3}+q(v)^{2}z^{4}.\label{Vaidyametriads2}\\
 A_v&=&2q(v)(1-z).\label{VaidyAv2}
\end{eqnarray}
It is easy to check when $v\rightarrow +\infty$, the above formula denotes a Vaidya-AdS-like metric \eqref{metriads1} while in the limit $v\rightarrow -\infty$, it corresponds to a AdS spacetime with nonvanshing $\beta$.
According to \cite{Caceres:2012em},   the external sources of current and energy-momentum tensor for the the above solution \eqref{Vaidyametriads1}-\eqref{VaidyAv2} are
\begin{eqnarray}
J^z_{(ext)}&=&2\frac{dq(v)}{dv},\\
T_{vv}^{(ext)}&=&z^2\frac{dm(v)}{dv}-2z^3q(v)\frac{dq(v)}{dv}.
\end{eqnarray}
It is noticed that  here in our solution, we consider the simple case that only the mass and charge depend on the time but the momentum relaxation coefficient $\beta$ does not vary with  time, which is similar as that  studied for five dimensional EMA theory \cite{Fernandez-Pendas:2019rkh}. This means that the momentum dissipation is treated more like as a probe because in this sense the background geometry is not obtained from solving Einstein-equations in a self-consistent way. More external sources  were considered to construct the solution in Vaidya setup of this theory \cite{Bagrov:2017tqn}, where the authors probed the  time-dependent optical conductivity without  translation invariance.

\section{Holographic thermalization}
For the purpose of exploring the dynamics and the scale dependence of the thermalization processes, the non-local observables are required to provide sufficient information. In this work, we will focus on two non-local observables, i.e., the equal time two-point function and entanglement entropy. Using the AdS/CFT correspondence, they can be calculated by the spacelike geodesic and the extremal volume in AdS space, respectively \cite{Balasubramanian:2011ur}. With the Vaidya-AdS black brane \eqref{dynamasspara}-\eqref{VaidyAv2} of the bulk theory in hands, we are ready to study the effect of momentum relaxation on this processes.
\subsection{Effect of momentum relaxation on two-point correlation functions}\label{sec:ads2point}
According to the study  in \cite{Balasubramanian:2011ur}, the on-shell equal time two-point correlation function for a scalar operator $\mathcal{O}$ with conformal dimension $\Delta$ can be holographic evaluated  as
\begin{equation}
\langle \mathcal{O}(t_0, x_i) \mathcal{O}(t_0, x_i')\rangle\sim \exp{(-\Delta\mathcal{ L})}
\end{equation}
where $\mathcal{ L}$ is length of  a bulk geodesic  connecting two  points $(t_0, x_i) $ and $(t_0, x_i')$ on the AdS boundary.  Thus, to disclose the properties of the equal time two-point correlation function, one  needs to minimize the length of $\mathcal{ L}$.

To proceed, we consider a space-like geodesic connecting the two boundary points: $(t, x_1) =
(t_0, -\ell/2)$ and $(t', x_1') = (t_0, \ell/2)$, with all other spatial directions identical at the two end points. $\ell$ is the boundary separation along $x_1\equiv x$. Such a geodesic then can be parametrized by $v = v(x)$ and $z = z(x)$ and the boundary conditions satisfied by this geodesic are
\begin{eqnarray}
z(\pm\ell/2)&=& z_0 , \quad v(\pm\ell/2)  = t_0, \label{2pointbouncondition1}\\
z(0)&=&z_*, \quad v(0)=v_*, \label{2pointbouncondition2}\\
z'(0)&=&v'(0) =0.\label{2pointbouncondition3}
\end{eqnarray}
where $z_0$ is the UV radial cut-off near the boundary; $z_*$ and $v_*$ are two parameters characterizing the extremal case that will be stated soon.
The length element of the geodesic is then given by
\begin{eqnarray}\label{ads2point1}
\mathcal{L}_{AdS}=\int^{\frac{\ell}{2}}_{-\frac{\ell}{2}}dx\frac{\sqrt{1-f(v,z)(v')^{2}-2v' z'}}{z},
\end{eqnarray}
where the prime denotes the derivative with respect to $x$. In order to evaluate the correlation functions, we need
to minimize the above length. It is obvious that the integral function in \eqref{ads2point1}, which can be treated as `Lagrangian', does not explicitly depend on $x$ which plays the role of time coordinate in classical
mechanics. Thus, the corresponding Hamiltonian is conservative with regard to $x$ and the related
conservation equation is
\begin{eqnarray}\label{2pointads}
1-f(v,z)\left(v'\right)^{2}-2 z' v'=\left(\frac{z_*}{z}\right)^{2}.
\end{eqnarray}
Subsequently, the equations of motion from the `Lagrangian' in \eqref{ads2point1} can be  reduced as
\begin{eqnarray}
z v''+2 z' v'-1+\left(v'\right)^{2}\left[f(v,z)-\frac{1}{2}z\frac{\partial f}{\partial z}\right]&=&0, \label{2pointmotion1}\\
z''+f(v,z)v''+z' v'\frac{\partial f}{\partial z}+\frac{1}{2}\left(v'\right)^{2}\frac{\partial f}{\partial v}&=&0. \label{2pointmotion2}
\end{eqnarray}

The minimal length of geodesic can be numerically calculated by solving the above equations of motion with the boundary conditions \eqref{2pointbouncondition1} and \eqref{2pointbouncondition2}. However, only the finite part of the geodesic length is physical and interested,  so one usually considers the renormalized length of the geodesic \cite{Balasubramanian:2011ur}
\begin{eqnarray}\label{renormalized2point}
\delta\mathcal{L}_{AdS}=\mathcal{L}_{AdS}+2 \ln{(z_0)},
\end{eqnarray}
where $2 \ln z_0$ is the contribution of a pure AdS boundary to eliminate the divergent term.

Before presenting the numerical results, we consider the ratio of the chemical potential and the temperature as a new varying parameter\footnote{Since the boundary field theory is conformal, it has been suggested that the only relevant parameter one can vary is this dimensionless ratio $\chi$ constructed from the chemical potential and the temperature if we consider the thermalization process with a chemical potential \cite{Caceres:2012em,Zhang:2015npb}. However, if we guarantee the existence of horizon and set $r_h=1$, one can easily show that varying $\chi$ is mathematically equal to varying $\beta$, as we will do later.}
\begin{eqnarray}\label{adsmuT}
\chi=\frac{1}{4\pi}\left(\frac{\mu}{T}\right)=\frac{2\,q_{0}\,z_{h}}{2+m_{0}\,z_{h}^{3}-2\,q_{0}^{2}\,z_{h}^{4}}.
\end{eqnarray}
Note the chemical potential $\mu$ defined by \eqref{adschemical} is dimensionless but the temperature has dimension  $T\sim [length]^{-1}$. So if we want to keep $\chi$ dimensionless we have to redefine $\mu$ by a scale with length unit that depends on the particular compactification. However, such a scale will not significantly change the behavior of $\chi$ but only as a factor. Besides, if we fix the mass parameter $m_0$ and the horizon radius $z_{h}$ then we can easily obtain a relation between $q_0$ and $\beta$ by \eqref{massandcharge} as well as  $T=0$,
\begin{eqnarray}\label{betacharge}
\beta^{2}=6\,z_{h}^{-2}-2\,q_{0}^{2}\,z_{h}^{2}.
\end{eqnarray}
Hereafter, for the sake of simplicity, we will fix $m_0=1$ and $z_h=1$. With such consideration, the maximal values of $q_0$ and $\beta$ are given by
\begin{eqnarray}\label{adsnumericalbetaq}
\beta^{2}=2\,q_{0}^{2},  \quad q_{0}^{2}\leq 3/2, \quad \beta^{2}\leq 3.
\end{eqnarray}
{Hence, $\chi$ is a monotonic function of $q_0$ or $\beta$ with range $[0, \infty]$. We therefore adopt $\beta$ as the varying parameter to proceed the numerical calculation.}

\begin{figure*}[]
\centering
\includegraphics[width=8cm]{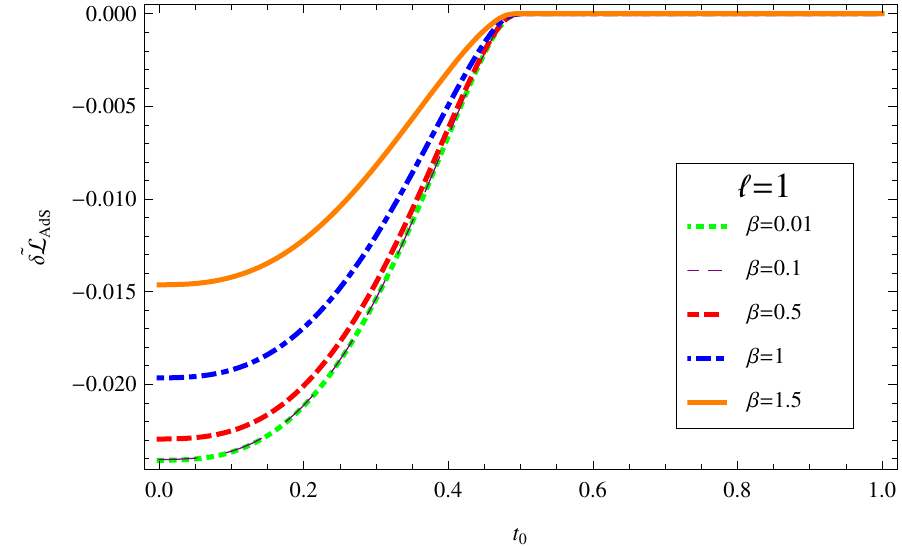}
\includegraphics[width=8cm]{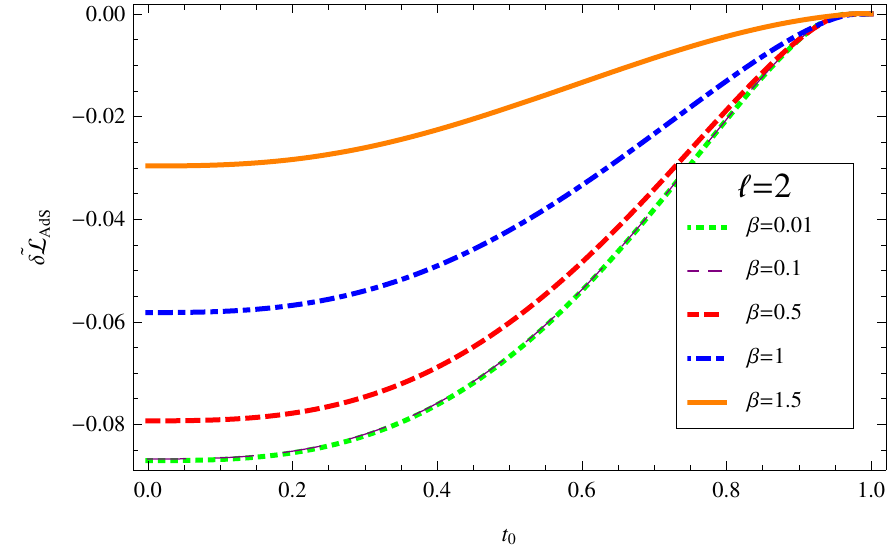}\\
\includegraphics[width=8cm]{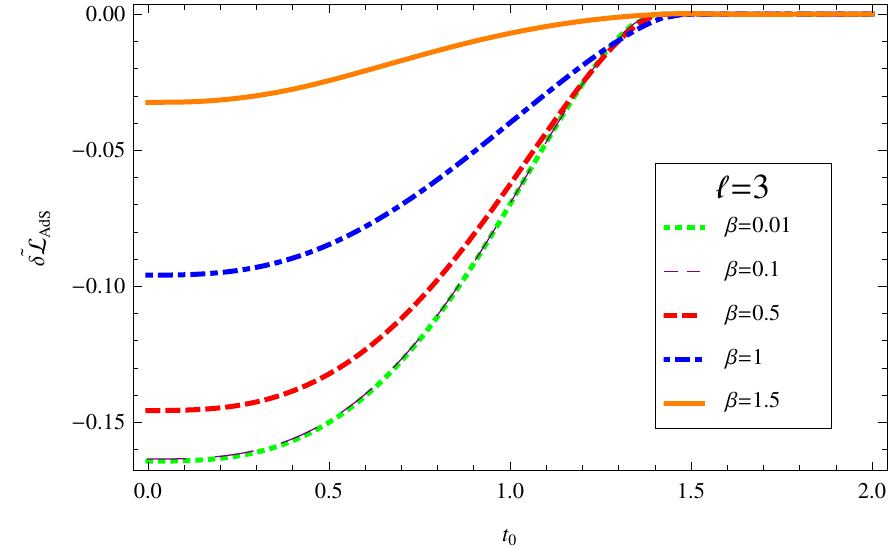}
\includegraphics[width=8cm]{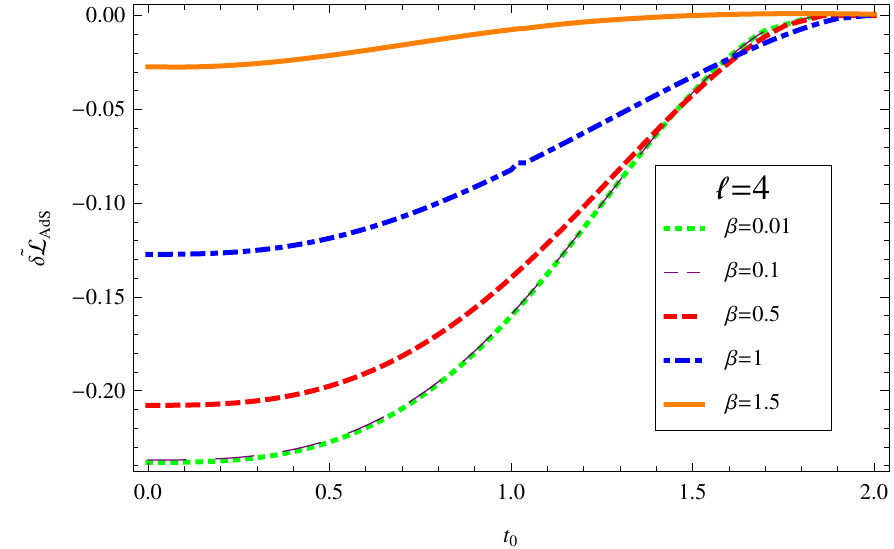}
\caption{The relative renormalized geodesic length $\tilde{\delta\mathcal{L}}_{AdS}$ as a function of boundary time $t_0$ with different $\beta$ for the boundary separation $\ell=1,\,2,\,3,\,4$. Clearly, for small $\beta$, the thermalization process is slightly effected by $\beta$.}
\label{fig:ads2point}
\end{figure*}

\begin{figure*}[]
  \includegraphics[width=8cm]{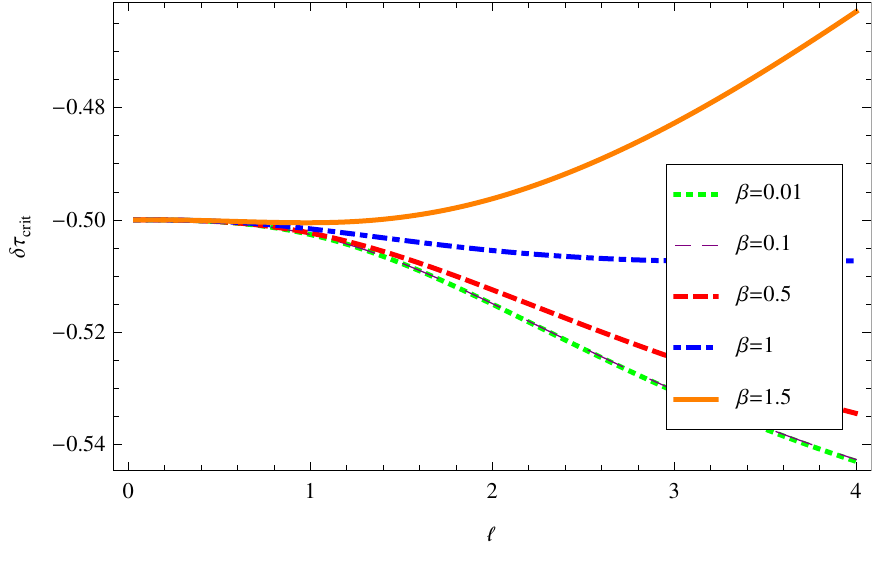}
  \includegraphics[width=8cm]{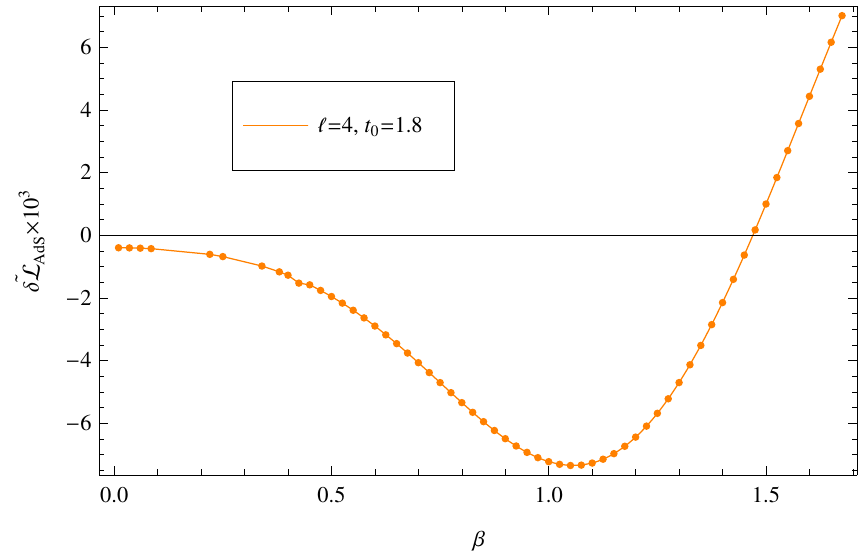}
  \caption{(Left panel) The relative critical thermalization time $\delta\tau_{crit}$ as a function of boundary separation $\ell$ with different $\beta$. (Right panel) The renormalized geodesic length $\tilde{\delta\mathcal{L}}_{AdS}$ as a function of linear coefficient $\beta$ with fixed boundary separation $\ell=4$ and time $t_{0}=1.8$.}
  \label{fig:ads2crittime}
\end{figure*}

In Fig.~\ref{fig:ads2point}, we show the relative renormalized geodesic length $\tilde{\delta\mathcal{L}}_{AdS}=\left(\delta\mathcal{L}_{AdS}-\delta\mathcal{L}_{AdS-thermal}\right)/\ell$ as a function of boundary time $t_0$ for different boundary separation $\ell$ and different $\beta$, where  $\delta\mathcal{L}_{AdS-thermal}$ denotes that the quantity is computed in static Vaidya-AdS spacetime.
The larger $\beta$ corresponds to larger renormalized geodesic. For small $\beta$ (for example, $\beta<0.1$), the length of geodesics are almost same. The figures also exhibit the tiny
dependence of the thermalization time $\tau_{real}$ on $\beta$ when $\ell$ is small, where $\tau_{real}$ is defined as the time when the probes reach the thermal equilibrium value. For large $\ell$, the curves show
a slightly complicated behavior for different $\beta$. The intersection of different curves near $\tau_{real}$ with large $\ell$ denotes the fact that different $\beta$ may lead to same renormalization geodesic, see the right panel of Fig.~\ref{fig:ads2crittime}. Further details will be discussed together with the results of holographic entanglement entropy.

Another interesting thermalization time is the so-called thermalization critical time $\tau_{crit}$, at which the tip of the extremal line or surface grazes the middle of the shell at $v=0$, defined as \cite{Li:2013cja}
\begin{equation}\label{ads2crittime}
\tau_{crit}=\int^{z_{*}}_{z_0}\frac{1}{f(z)}dz,
\end{equation}
where $f(z)=1-\frac{1}{2}\beta^{2}z^{2}-m_{0}z^{3}+q_{0}^{2}z^{4}$. The left panel of Fig.~\ref{fig:ads2crittime} shows the relation between the relative critical thermalization time $\delta\tau_{crit}$  and $\ell$, where  $\delta\tau_{crit}$ is re-scaled as
\begin{eqnarray}\label{ads2pointrelative}
\delta\tau_{crit}=\frac{\tau_{crit}-\ell}{\ell}.
\end{eqnarray}
The numerical results show several interesting properties. First, $\tau_{crit}$ is also nearly independent of the coefficient $\beta$ when $\ell$ is small, and appears as a constant. In other words, $\tau_{crit}$ saturates the causality bound $\tau_{crit} \sim \frac{\ell}{2}$ \cite{Li:2013cja,AbajoArrastia:2010yt}. This feature can be understood easily, since when the boundary separation is small the metric could be seen as an asymptotically AdS, and the geodesic in the static black brane will only extend near the boundary. Second, the critical thermalization time has monotonic dependence on $\beta$, i.e., $\tau_{crit}$ increases as $\beta$ increases, when $\ell$ is large enough. This is similar with $\tau_{real}$ from the right-bottom panel of Fig.~\ref{fig:ads2point} and we has numerically checked that $\tau_{crit}\approx\tau_{real}$ in our case.

\subsection{Effect of momentum relaxation on holographic entanglement entropy}\label{sec:adsHEE}

We now consider the case of holographic entanglement entropy which is modeled by the extremal surfaces of a spherical region in the boundary. The prescription for computing entanglement entropy using the AdS/CFT correspondence has been initially proposed in \cite{Ryu:2006bv,Nishioka:2009un}, in which it was addressed that for a system $A$ in  the boundary  CFT  which has a gravity dual,  the information included in a subsystem $B$ is
evaluated by the entanglement entropy $S_A$ as
 \begin{equation}\label{RTF}
S_A=\frac{\mbox{Area}(\gamma_A)}{4G^{(d+2)}_N}\ .
\end{equation}
where $\gamma_A$ is a  $d$-dimensional minimal surface with boundary given by the $(d-1)$-dimensional manifold
$\partial \gamma_A=\partial A$, and  $G^{(d+2)}_N$ is the Newton constant of the general gravity in AdS$_{d+2}$ theory.

Since we are working in a four dimensional Vaidya-AdS spacetime, the spherical region in the boundary is actually a disk, and the computation of the entanglement entropy is then identical to the Wilson loop computation. Thus, we will use the polar coordinates $(\rho, \phi)$ to rewrite the disk. The extremal surface then can be represented by $z(\rho)$ and $v(\rho)$ because of the azimuthal symmetry in the $\phi$-direction.  Subsequently,
the volume element  of hyper-surfaces which measures the entanglement entropy in our model  is given by
\begin{equation}\label{adsEE}
\mathcal{V}_{AdS}\equiv\mbox{Area}(\gamma_A)=\int^{R}_{0}d\rho\,\frac{\rho}{z^2}\sqrt{1-f(v, z)\,(v')^{2}-2\,v'\,z'},
\end{equation}
where $'\equiv\frac{d}{d\rho}$, and we have absorbed a factor $(2\pi)^{-1}$ to $\mathcal{V}_{AdS}$. The equations of motion derived from \eqref{adsEE} are
\begin{widetext}
\begin{eqnarray}
z''&=&\frac{1}{2\rho z}\Big[4\rho f^{2} (v')^{2}+f\left(8\rho v' z'-4\rho+z (v')^{2}(2z'-\rho\frac{\partial f}{\partial z})\right)\nonumber\\&&+z\left(4v'(z')^{2}-2z'(1+\rho v'\frac{\partial f}{\partial z})-\rho (v')^{2}\frac{\partial f}{\partial v}\right)\Big], \label{adseemotion1} \\
v''&=&\frac{1}{2\rho z}\left[4\rho\left(1-(v')^{2}f-2v'z'\right)+zv'\left(2(v')^{2}f-2+v'(4z'+\rho\frac{\partial f}{\partial z})\right)\right].\label{adseemotion2}
\end{eqnarray}
\end{widetext}

 Different with the case of the two-point function, here the integral function or Lagrangian of \eqref{adsEE}  implicitly depends on $\rho$, which means there is no conservation equation. Meanwhile, we cannot construct the solutions from the midpoint to impose the boundary conditions because $\rho$ in the denominator will cause numerical issues at $\rho=0$. Therefore we expand the equations of motion at a point $\rho_{p}$ near $\rho=0$ to quadratic order, and the results will fix the boundary conditions at this point as
\begin{eqnarray}\label{adseeboundary1}
z(\rho_{p})=z_{*}-\frac{f(v_{*},z_{*})}{2z_{*}}\rho_{p}^{2},\quad v(\rho_{p})=v_{*}+\frac{\rho_{p}^{2}}{2z_{*}},
\end{eqnarray}
where $z_{*}, v_{*}$ are two free parameters, and $z'(\rho_{p}), v'(\rho_{p})$ can also be set by this expansion. The other two boundary conditions read
\begin{eqnarray}\label{adseeboundary2}
z(R)=z_0,\quad v(R)=t_0,
\end{eqnarray}
where again, $z_0$ is the radial IR cut-off and $t_0$ is the boundary time.

Then we can solve the equations of motion \eqref{adseemotion1} and \eqref{adseemotion2}  via the boundary conditions \eqref{adseeboundary1} and \eqref{adseeboundary2}. During the calculation, we  renormalize the  AdS volume by subtracting the cut-off dependent part \cite{Balasubramanian:2011ur}
\begin{eqnarray}\label{renormalizedhee}
\delta\mathcal{V}_{AdS}=\mathcal{V}_{AdS}-\frac{R}{z_0}.
\end{eqnarray}
\begin{figure*}[htbp]
\centering
\includegraphics[width=8cm]{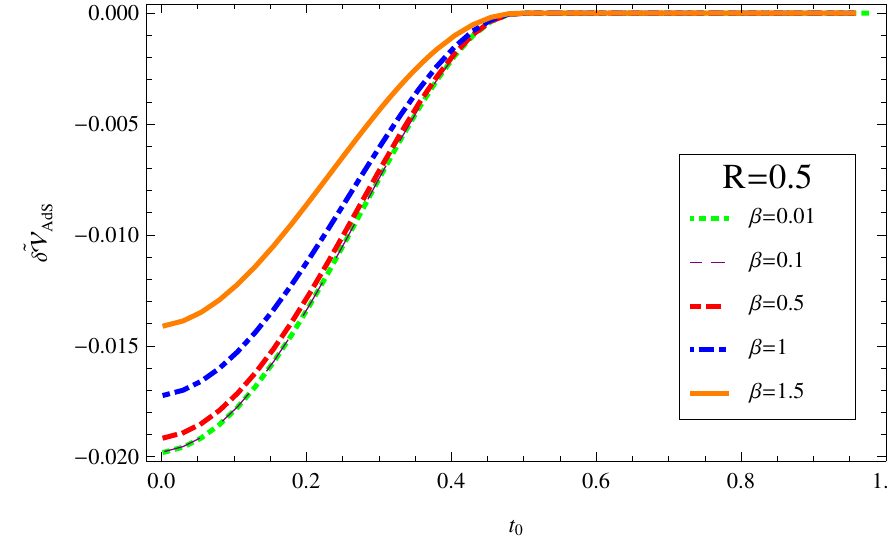}
\includegraphics[width=8cm]{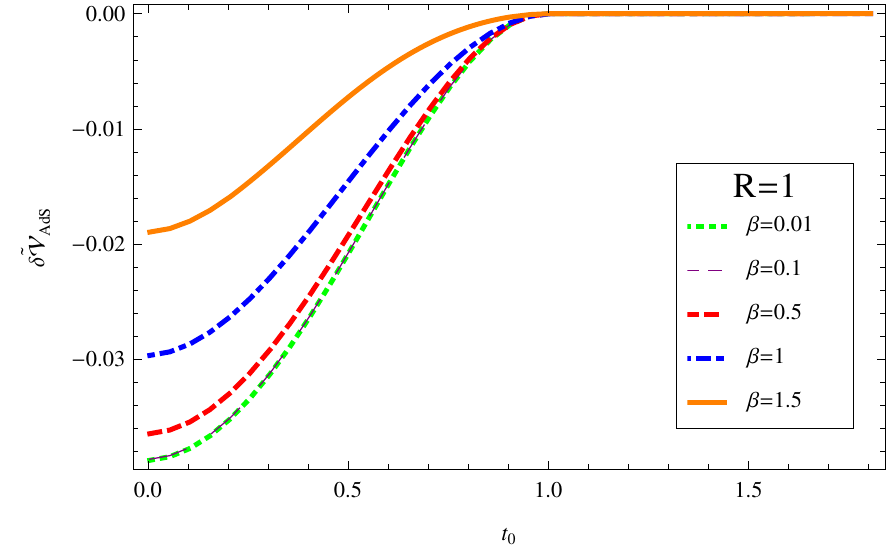}\\
\includegraphics[width=8cm]{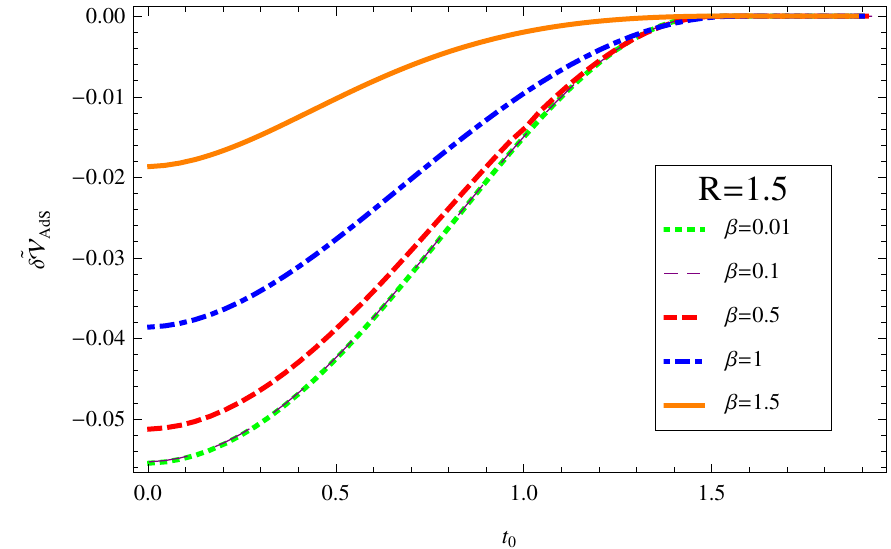}
\includegraphics[width=8cm]{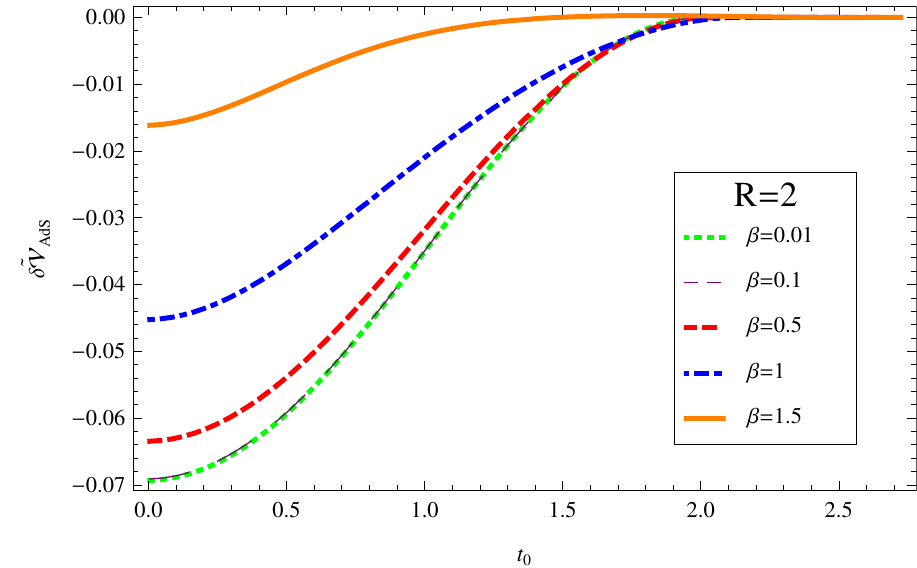}
\caption{The relative renormalized volume element $\tilde{\delta\mathcal{V}}_{AdS}$ as a function of boundary time $t_0$ with different $\beta$ for the boundary radius $R=0.5,\,1,\,1.5,\,2$.}
\label{fig:adsee}
\end{figure*}
The results of $\tilde{\delta\mathcal{V}}_{AdS}=\left(\delta\mathcal{V}_{AdS}-\delta\mathcal{V}_{AdS-thermal}\right)/(\pi R^{2})$ against $t_0$ are shown in Fig.~\ref{fig:adsee}. It is obvious that the features of entanglement entropy during the thermalization processes are very similar with those in geodesic case. The larger $\beta$  gives  larger renormalized entanglement entropy while again for small $\beta$ they are almost same. Moreover, when $R$ is small,  the thermalization time $\tau_{real}$ is almost the same for different $\beta$, while for large $R$ the time $\tau_{real}$ behaves as a multi-valued function of $\beta$ as we observed in the two-point function case. This means that  different $\beta$ may lead to same renormalized entropy, see the right panel of Fig.~\ref{fig:adseecrittime}.

The relative critical time $\delta\tau_{crit}$ as a function of the boundary radius $R$ has been presented in the left panel of Fig.~\ref{fig:adseecrittime}. For small $\beta$ and $R$ the critical time is linear with $R$, but for large $\beta$ the linearity does not hold. In particular, the critical time $\tau_{crit}$ violates the causal bound, i.e., $\tau_{crit}>R$. It is also obvious when $R$ is large enough, the critical thermalization time increase monotonically as $\beta$ increases, which matches the observe obtained from two-point correlation function. This suppression effect of momentum relaxation on the thermalization process may be explained as follows. In holography, the approach to thermal equilibrium in the boundary field theory is described by the decay of the bulk fluctuations. And it was found in \cite{Kuang:2017cgt} that the
shift of the quasi-normal frequencies from zero for the bulk fluctuations is larger if the system has bigger $\beta$.
However, the deep physical understanding of this suppression from the field theory is still missing.
\begin{figure*}[]
\centering
  \includegraphics[width=8cm]{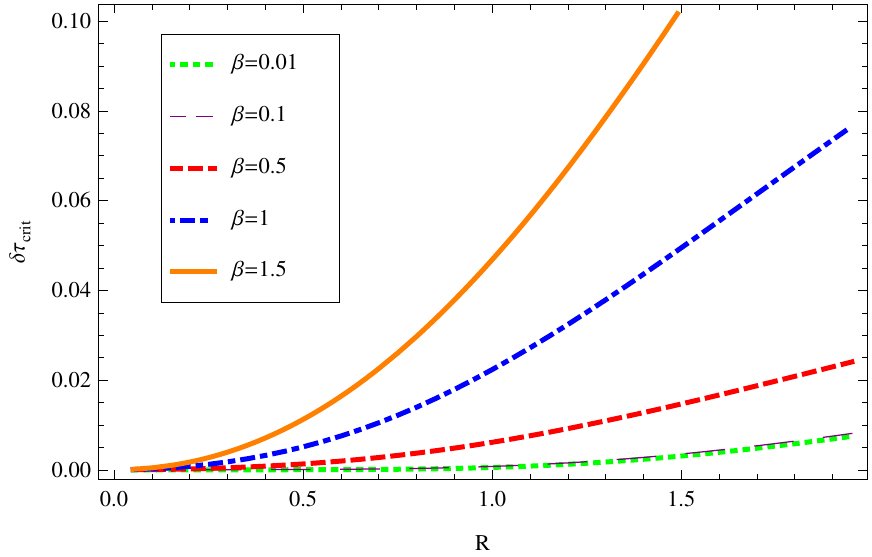}
  \includegraphics[width=8cm]{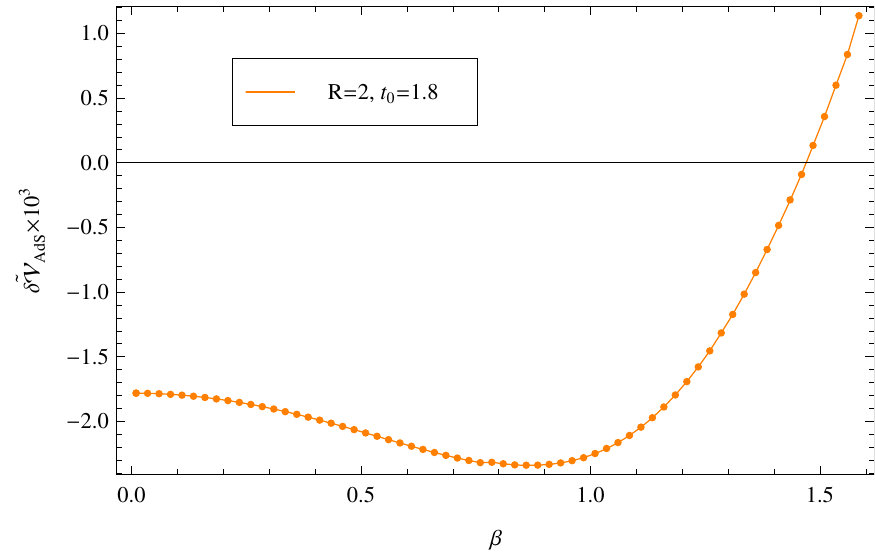}
  \caption{(Left panel) The relative critical thermalization time $\delta\tau_{crit}$ as a function of boundary radius $R$ with different $\beta$. (Right panel) The relative renormalized volume $\tilde{\delta\mathcal{V}}_{AdS}$ as a function of linear coefficient $\beta$ with fixed boundary radius $R=2$ and time $t_{0}=1.8$.}
  \label{fig:adseecrittime}
\end{figure*}
\begin{figure*}
  \includegraphics[width=8cm]{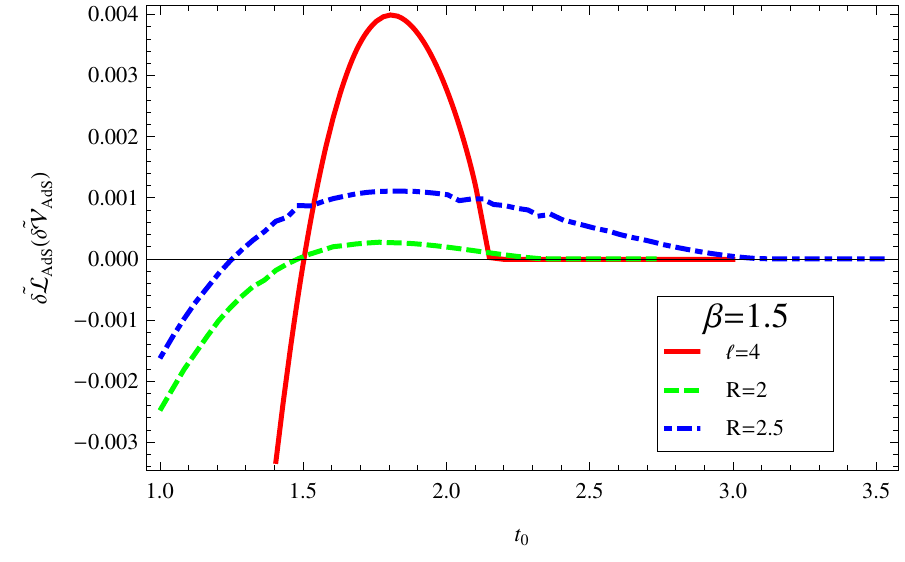}
  \includegraphics[width=7.6cm]{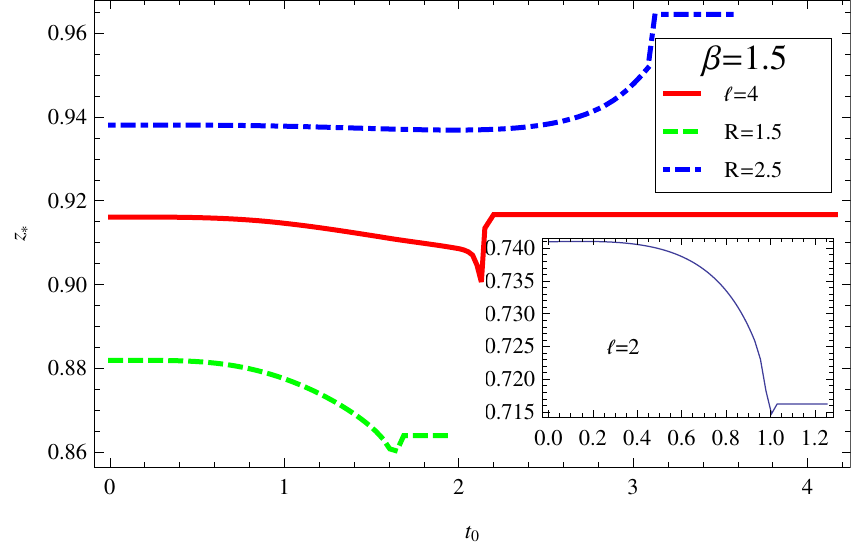}
  \caption{(Left panel) The relative renormalized quantities $\tilde{\delta\mathcal{L}}_{AdS}$ or $\tilde{\delta\mathcal{V}}_{AdS}$ as a function of boundary separation scale with $\beta=1.5$ for $\ell=4$, $R=2$ and $R=2.5$. (Right panel) $z_{*}$ as a function of $t_0$ for $\beta=1.5$ with boundary separation scale $\ell=4$, $R=1.5$ and $R=2.5$. The embedded picture shows the case when $\ell=2$. Note here the dashed green line means different $R$ because for $R=1.5$ its behavior is close to zero when approaching the thermal equilibrium point.}
  \label{fig:adstime2}
\end{figure*}

In the right panel of Fig.~\ref{fig:adseecrittime}, we observe similar but weaker feature compared to the geodesic case that for fixed large $R$, different $\beta$ leads to same $\tilde{\delta\mathcal{V}}_{AdS}$  if $t_0$ is in a certain range. The thermalization process shows such time range is usually a short interval before it reaches the thermal equilibrium. For example, see the right-bottom panel of Fig.~\ref{fig:adsee}, the interval is around $1.6<t_0 <2.0$. With smaller boundary separation leads to narrower interval. It can be expected that for small enough $\ell \,(R)$ the renormalized $\tilde{\delta\mathcal{L}}_{AdS} (\tilde{\delta\mathcal{V}}_{AdS})$ will be a monotonic function of $\beta$. Taking care of the two observables, we find that the right panels of Fig.~\ref{fig:ads2crittime} and Fig.~\ref{fig:adseecrittime} illustrate another interesting property. The relative renormalized geodesic length $\tilde{\delta\mathcal{L}}_{AdS}$( $\tilde{\delta\mathcal{V}}_{AdS}$) becomes positive for fixed large boundary separation (radius, $\ell=4$ or $R=2$) and time when $\beta \gtrsim 1.47$ for both probes.
However, this feature is not hold for small enough $\ell$ or $R$. For clarity, we zoom in the thermalization curves near the thermal equilibrium point, and plot the relation between $z_{*}$ and $t_0$. In details, from the left panel of  Fig.~\ref{fig:adstime2}, we see that for two-point function when $\ell=4$ the thermalization process will cross the zero point and then rapidly return to thermal equilibrium\footnote{Here `rapid' behavior is described in the sense that the time interval from the maximum $\tilde{\delta\mathcal{L}}_{AdS}(t_0\simeq1.84)$ to the equilibrium is only about $0.34$, while the thermalization time $\tau_{real}\simeq2.18$.}, while for holographic entanglement entropy the process is quite sluggish. These behaviors are similar with the evolution of holographic entanglement entropy and holographic complexity in the massive BTZ black hole, which also states the appearance of momentum relaxations in the boundary CFT theory \cite{Yuting:2019prd}. The right panel of Fig.~\ref{fig:adstime2} shows that $z_*$ in equilibrium is larger than its initial value if the boundary separation and $\beta$ are both large enough. When the shell is approaching the extremal line vertex, $z_*$ will jump to its equilibrium value. Besides, the different evolving behaviours of $\tilde{\delta\mathcal{L}}_{AdS}$ and $\tilde{\delta\mathcal{V}}_{AdS}$ correspond to different jumping effects for large $\ell$ and $R$, see the red and dot-dashed blue lines. In contrast, we note that for small $\ell$ or $R$ the thermalization process of two observers are similar, see the green line and the inserted picture in the right panel of Fig.~\ref{fig:adstime2}. We claim that these positive-features result from the effect of momentum relaxation. If $\beta=0$ these features will disappear, for example, see figure 10 and 23 in \cite{Caceres:2012em}. {It is noticed that the universal properties of HEE holographic thermalization has been analytically studied in \cite{Liu:2013iza,Liu:2013qca}, but their analysis should be carefully generalized to our case because of the appearence of momentum relaxation. It would be interesting to develop an analytical method to find the
the expressions of $z_{*}(t_0)$ and $\tilde{\delta\mathcal{L}}_{AdS}(t_0)(\tilde{\delta\mathcal{V}}_{AdS}(t_0))$, which we hope to address in the future.}

\begin{figure*}[htbp]
\centering
\includegraphics[width=8cm]{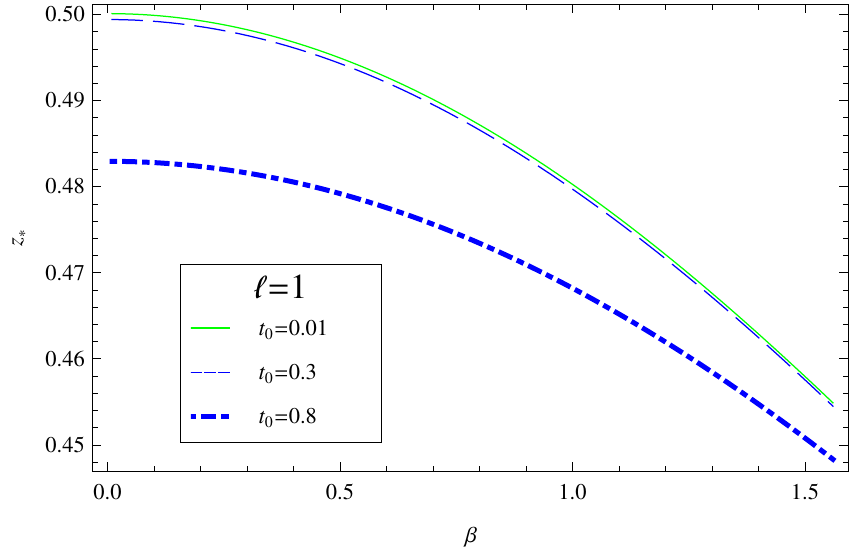}
\includegraphics[width=8cm]{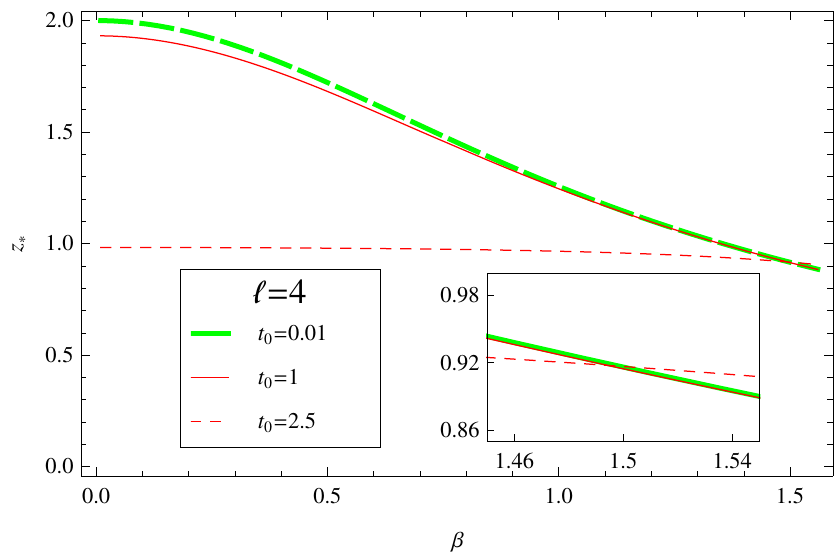}\\
\includegraphics[width=8cm]{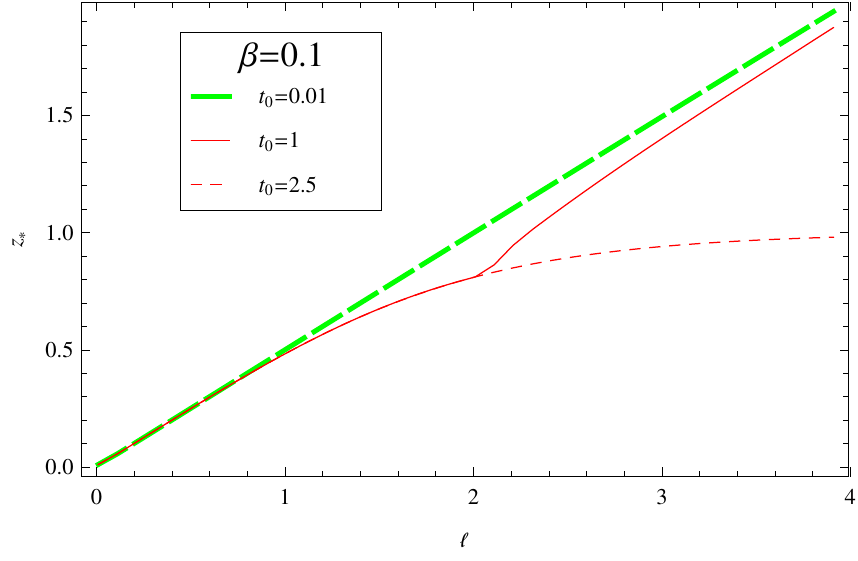}
\includegraphics[width=8cm]{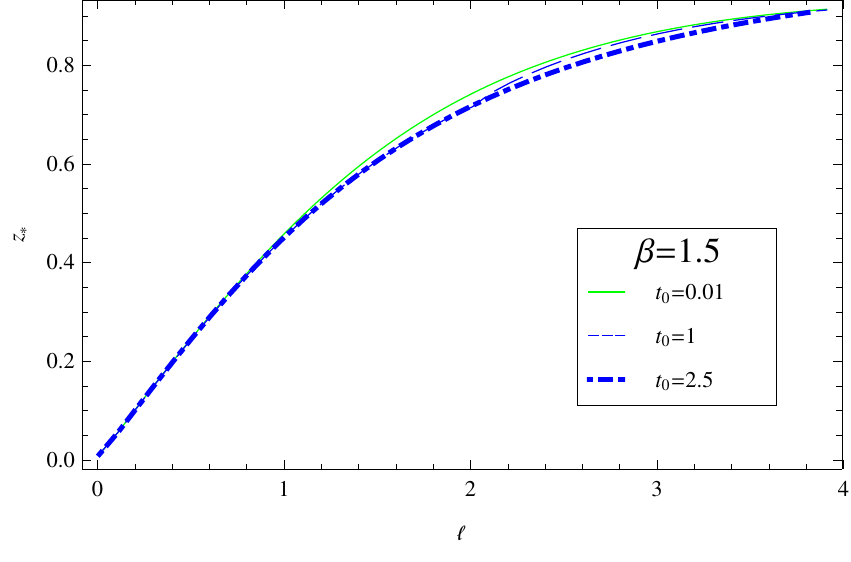}
\caption{(Top) The relation between $z_*$ and $\beta$ with $\ell=1,\, t_0=0.01,\, 0.3,\, 0.8$ and $\ell=4,\, t_0=0.01,\, 1,\, 2.5$, respectively. (Bottom) The relation between $z_*$ and $\ell$ with $\beta=0.1,\,1.5$ at fixed boundary time $t_0=0.01,\,1,\,2.5$, respectively.}
\label{fig:ads2pointbetaLzs}
\end{figure*}

\begin{figure*}[htbp]
\centering
\includegraphics[width=8cm]{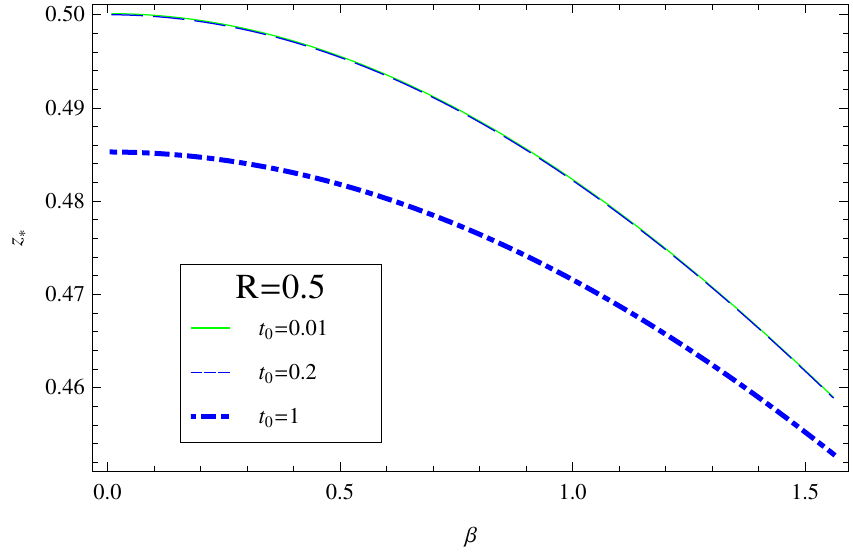}
\includegraphics[width=8cm]{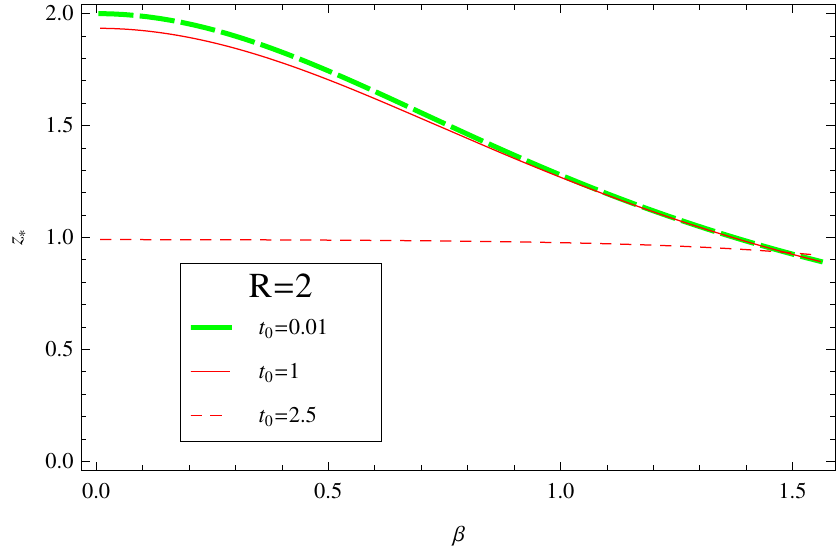}\\
\includegraphics[width=8cm]{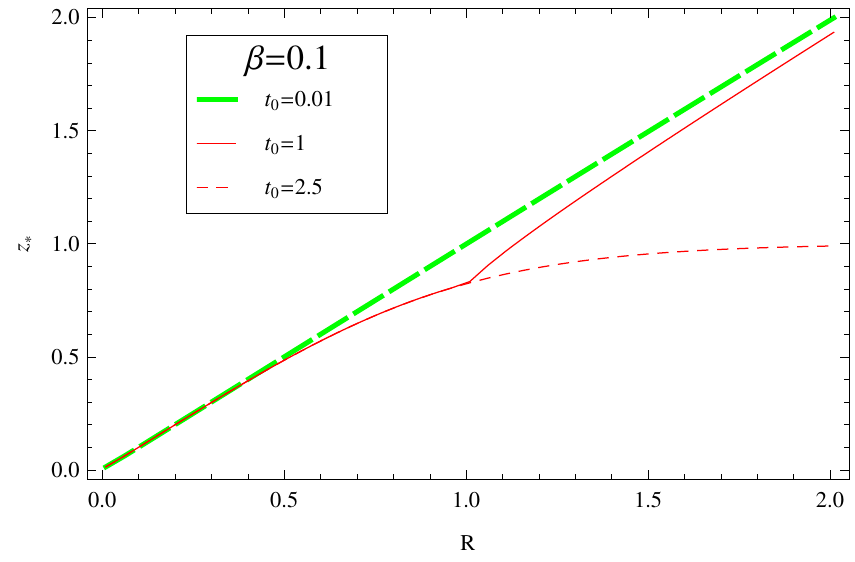}
\includegraphics[width=8cm]{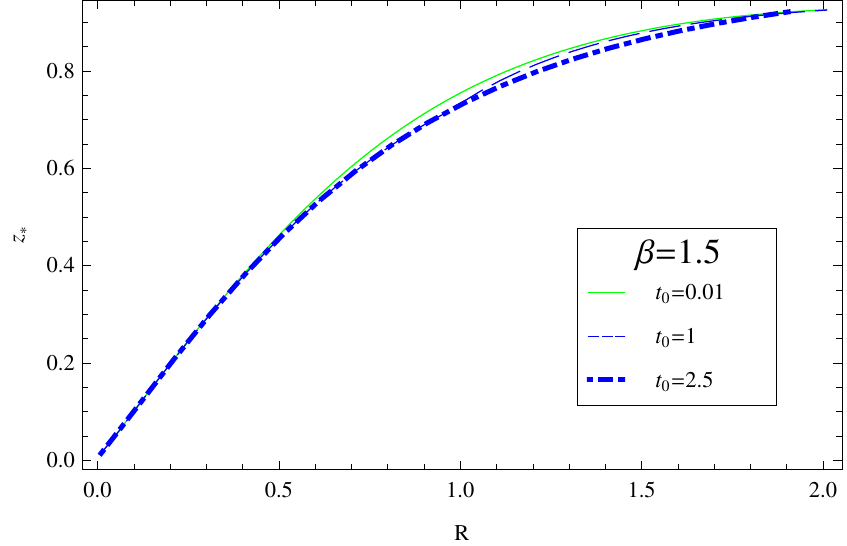}
\caption{(Top) The relation between $z_*$ and $\beta$ with $R=0.5,\, t_0=0.01,\, 0.2,\, 1$ and $R=2,\, t_0=0.01,\, 1,\, 2.5$, respectively. (Bottom) The relation between $z_*$ and $R$ with $\beta=0.1,\,1.5$ at fixed boundary time $t_0=0.01,\,1,\,2.5$, respectively.}
\label{fig:adseebetaRzs}
\end{figure*}

We also note that when the boundary separation is large enough, the equilibrium value of $z_{*}$ is larger than the non-equilibrium value. Thus it will be interesting to present the relation of $z_*$ with $\beta$ or the boundary separation at fixed time for two probes, see Fig.~\ref{fig:ads2pointbetaLzs} and Fig.~\ref{fig:adseebetaRzs}. Clearly the corresponding curves of two probes behave similarly. So we can focus on the two-point function to discuss these behaviors.

The two top panels of Fig.~\ref{fig:ads2pointbetaLzs} show the behaviors of $z_*$ as a function of $\beta$ at three time points for two boundary separations. These three time points are chosen as the initial time, one arbitrary non-equilibrium point-in-time and one point-in-time after equilibrium. For large separation $\ell=4$, we chose $t_0=0.01,\, 1,\,2.5$ and for $\ell=1$, $t_0=0.01,\, 0.3,\,0.8$. The figures illustrate that the thermal equilibrium value of $z_*$ is always smaller than $1$. While for small enough $\ell$, $z_*<1$ is true for entire thermalization process. Besides, we can tell that for large enough $\ell$ the intersection of $z_*$ at different time solely happens around $\beta\approx1.5$, which means the case that the equilibrium value of $z_*$ is larger than non-equilibrium $z_*$ can only be true at large $\beta$.

The two bottom panels of Fig.~\ref{fig:ads2pointbetaLzs} present the behavior of $z_*$ as a function of $\ell$ for $\beta=0.1,\,1.5$ at $t_0=0.01,\, 1,\,2.5$. For left panel, $\beta$ is small enough. The green dashed line means the thermalization process just begins for the boundary separation region we chose. The spacetime is a pure AdS-like geometry with very small $\beta$, thus the tips of extremal line (i.e., $z_*$) are following $z _{*}\sim\ell/2$ (the extremal line behaves like a half circle.) The red dashed line means the equilibrium has been reached by all $\ell$ in the region. This can be checked again from the right bottom panel of Fig.~\ref{fig:ads2point}. The spacetime then behaves as a Vaidya-AdS type geometry with very small $\beta$. The tips of the geodesic are smaller than the green dashed line case for large $\ell$ but almost same for small enough $\ell$, as expected. Moreover, Fig.~\ref{fig:ads2crittime} shows that $\ell=4$ will have largest $\tau_{crit}<2.5$, which we denote it as $\tau^{max}_{crit}$. Any $t_0$ in the region \{0, $\tau^{max}_{crit}$\} will relate to an unique $\ell$, which the thermalization time corresponding to this unique boundary separation is $t_0$. We further denote it as $\ell_{lif}(t_0)$. Then for fixed $t_0$, all boundary separations that are smaller than $\ell_{lif}$ correspond to the equilibrium states, while these greater than $\ell_{lif}$ ones are in non-equilibrium states, see the red solid line.

For right panel where $\beta$ is large, the differences of $z_*$ curves at different time are weak. The comparison between two panels suggests the effect of $\beta$ on the thermalization process is significant.

\section{Conclusion and discussion}
In this paper, we investigated  the holographic thermalization process  in a four dimensional Einstein-Maxwell-axions gravity theory, which is  dual to a boundary theory with momentum relaxation.
We mainly calculated  the equal time two-point correlation function of a scalar operator and  entanglement entropy, which are modeled by the minimal lengths of geodesic  and volumes of spherical region in AdS space. We focused on the effects of momentum relaxation  on this process. It was found that  the momentum relaxation would suppress the holographic
thermalization process from the behaviors of two probes. This result is different from that found in massive gravity \cite{Hu:2016mym}. It may be because the thermal equilibrium configurations in two models are different, or here the momentum relaxation is introduced through new matter fields so that new degree of freedom is involved.  Or we may extend our studies in a generalized  axion theory with higher order(see \cite{Alberte:2015jhep}) to consider a generic potential and investigate the thermalization process.
Deep physical explanation  deserves more efforts.

We argued that {the suppression} effect on the thermalization process can be explained by profound violation of the quasi-normal frequencies from zero mode  for the bulk fluctuations studied in \cite{Kuang:2017cgt}, since the decay of the bulk fluctuations holographically describes the approach to thermal equilibrium in the dual boundary field theory. We believe that the deep physical insight of this {suppression} in the viewpoint of the field theory is encouraged to be investigated.

Here we chose the circle geometry to evaluate the holographic entanglement entropy. It is straightforward to study it by the method of stripe geometry with the same strategy. Moreover, it was addressed in \cite{Susskind:2014moa} that the entanglement entropy
is usually not enough to describe the rich geometric structure because it grows in a very
short time during the thermalization process of a strongly coupled system.  Recently, another
active studies in holographic framework is the holographic complexity evaluated from gravity side via `complexity=volume' conjecture \cite{Alishahiha:2015rta}
or  `complexity=action' conjecture \cite{Brown:2015bva,Brown:2015lvg}.
The study on the evolution of sub-region complexity may help us supplement the
description of thermalization process. This proposal has been addressed for Einstein gravity in \cite{Chen:2018mcc} and recently was generalized in \cite{Ling:2018xpc,Zhang:2019vgl,Zhou:2019jlh}. Thus, it would be
very interesting to study the effect of momentum relaxation on the evolution of complexity during the thermalization process in this model,
which will be presented elsewhere.

\section*{Acknowledgments}
We appreciate Matteo Baggioli and Shang-Yu Wu for helpful corresponding. This work is supported by the Natural Science Foundation of China under Grant No.11705161, Fok Ying Tung Education Foundation under Grant No. 171006 and Natural Science Foundation of Jiangsu Province under Grant No.BK20170481.

\end{document}